\documentclass[aps,pre,longbibliography,notitlepage,floatfix,11pt,tightenlines,nofootinbib]{revtex4-1}

\usepackage{enumitem}

\usepackage{bm,amsmath,amssymb,graphicx,stmaryrd}
\graphicspath{ {Figures/} }
\usepackage[abs]{overpic}

\usepackage[format=plain,labelfont={bf,small},textfont=small,justification=raggedright,singlelinecheck=false]{caption}

\begin{document}

\title{The saturation bifurcation in coupled oscillators}
\author{H. G. Wood}
\author{A. Roman}
\affiliation{Department of Biomedical Engineering and Mechanics}
\author{J. A. Hanna}
\email{hannaj@vt.edu}
\affiliation{Department of Biomedical Engineering and Mechanics, Department of Physics, Center for Soft Matter and Biological Physics, Virginia Polytechnic Institute and State University, Blacksburg, VA 24061, U.S.A.}

\date{\today}

\begin{abstract}
We examine examples of weakly nonlinear systems whose steady states undergo a bifurcation with increasing forcing, such that a forced subsystem abruptly ceases to absorb additional energy, instead diverting it into an initially quiescent, unforced subsystem.
We derive and numerically verify analytical predictions for the existence and behavior of such saturated states for a class of oscillator pairs.  We also examine related phenomena, including zero-frequency response to periodic forcing, Hopf bifurcations to quasiperiodicity, and bifurcations to periodic behavior with multiple frequencies.
\end{abstract}

\maketitle

\section{Introduction}

Coupled weakly nonlinear oscillators underly much analysis of pattern formation and vibrations, and often serve as reduced models of continuous systems such as fluid flows.
In this note, we explore a remarkable effect present in some coupled oscillator systems with internal resonances. In such autoparametric systems, it is possible to resonantly force an oscillator, yet have its steady-state amplitude abruptly saturate at a value independent of the forcing amplitude, with all the additional energy of forcing transferred to another, unforced, oscillator coupled to the first through a harmonic resonance.
First observed in the 1970s in a mathematical toy model of ship motions \cite{Nayfeh73, NayfehMook78}, this \emph{saturation} phenomenon has been experimentally confirmed in related experimental models \cite{Oh00} and simple compound structures \cite{Haddow84, NayfehZavodney88, BalachandranNayfeh91}.
However, the saturation phenomenon is surprisingly little known in the broader world of physics and engineering, and to date no systematic study has been performed to understand the origins and predict the existence of the effect for a given nonlinearity, even in the simplest case of two oscillators.  
Here we derive and demonstrate the effectiveness of an analytical criterion for prediction of the saturation bifurcation in a class of coupled oscillator pairs with direct and mixed nonlinearities of various orders. 
 We show that saturation may occur for several types of nonlinearity if certain conditions are met, 
 and explain the hitherto unremarked zero-frequency (DC) response of some saturated systems, including the original ship example \cite{Nayfeh73}.  We also examine additional effects, such as the loss of stability of some saturated solutions through a Hopf bifurcation to quasiperiodic response
 , and the existence of an alternate bifurcation to multi-frequency periodic solutions for some systems that do not experience saturation. 

The saturation effect may prove useful in the design of materials or structures that absorb \cite{HaxtonBarr72, Vakakis01}, localize \cite{Hodges82, FilocheMayboroda09, Cuevas09}, or divert \cite{Alam12} vibrational or wave energy.
The effect can also appear in weakly nonlinear approximations of driven hanging cables \cite{TadjbakhshWang90, LeePerkins92} or spring-pendulum systems \cite{LeeHsu94}, in which the relevant couplings contain time derivatives.

\section{Formulation and Solution}

Consider the pair of coupled oscillators 
\begin{align}
\ddot \theta + \omega_\theta^2\theta &= -c_\theta\dot\theta - k_\theta\phi^m\theta^n + f\cos\left(\omega_\theta t\right) \, , \label{thetaosc} \\
\ddot \phi + \left(r\omega_\theta\right)^2\phi &= -c_\phi\dot\phi - k_\phi\phi^l\theta^q \, , \label{phiosc}
\end{align}
where the $c$'s, $k$'s, and $f$ are positive constants of order $\epsilon \ll  1$, $l$, $m$, $n$, and $q$ are integers, and $r$ is a rational number.  
For clarity and simplicity, we leave detuning parameters out of the analysis,\footnote{In a curious paper, Oueini and co-workers \cite{Oueini99} examine a particular system with direct internal resonance ($r=1$) and some combination of cubic Duffing-type and coupling nonlinearities containing time derivatives, and show that detunings appear to create a pair of saturated states.  Whether such behavior is unique or common remains an open question.   
Incidentally this appears to be the only other paper, besides the present one, to demonstrate saturation in systems with cubic nonlinearities; early literature often repeats the misconception that such systems do not saturate.} and ignore the distinction between natural and resonant frequencies, which are quite close. 
Thus, we think of the $\theta$ oscillator as resonantly forced, and coupled via a sub- or super-harmonic internal resonance with the unforced oscillator $\phi$.
While more general expressions than (\ref{thetaosc}-\ref{phiosc}) can be written, this form will allow us to easily display the important features of saturating systems, and to look at the effects of particular polynomial nonlinearities in isolation.  General polynomial terms may arise, or dominate, for any number of reasons in conservative or nonconservative physical systems, an issue which we do not explore here.  
A subset of our coupling terms may come from a term in a potential of the form $\phi^m\theta^q$ with $m = l + 1$ and $q = n + 1$.


Any system like (\ref{thetaosc}-\ref{phiosc}) with $l \ne 0$ can display a trivial response where $\phi$ is quiescent and $\theta$ is a forced, damped, linear oscillator with a steady-state amplitude linear in the forcing $f$.  The $l=0$ cases will be discussed separately below.
An oscillator pair undergoing saturation will have a second fixed point for which the forced amplitude is constant and the unforced amplitude of $\phi$ depends on the forcing applied to $\theta$.

We solve the system (\ref{thetaosc}-\ref{phiosc}) perturbatively using a variant of the method of averaging, employing
 the \emph{ansatz} $\theta = \alpha_\theta\cos\left(\omega_\theta t + \beta_\theta\right), \phi = \alpha_\phi\cos\left(r\omega_\theta t + \beta_\phi\right)$, where the $\alpha$'s and $\beta$'s are slowly varying functions of time.
The zeroth order system is two uncoupled oscillators.  At the next order, we balance single time derivatives and the leading order of the projection of the right hand sides of (\ref{thetaosc}-\ref{phiosc}) onto the resonant frequencies of the zeroth order solution.\footnote{Several approaches to deriving averaged equations are discussed in texts such as \cite{NayfehMook79, GuckenheimerHolmes83, Strogatz94}.  That described here is closest to one of the two presented in \cite{Strogatz94}.}
The slow equations for the amplitudes and phases derived in this manner
 have the general form
\begin{align}
\dot\alpha_\theta &= -\frac{c_\theta\alpha_\theta}{2} - \frac{f}{2\omega_\theta}\sin\beta_\theta + k_\theta\alpha_\phi^m\alpha_\theta^n I_{\alpha\theta} \, , \label{amptheta} \\
\alpha_\theta\dot\beta_\theta &= -\frac{f}{2\omega_\theta}\cos\beta_\theta + k_\theta\alpha_\phi^m\alpha_\theta^n I_{\beta\theta} \, , \label{phasetheta}\\
\dot\alpha_\phi &= -\frac{c_\phi\alpha_\phi}{2} + k_\phi\alpha_\phi^l\alpha_\theta^q I_{\alpha\phi} \, , \label{ampphi} \\
\alpha_\phi\dot\beta_\phi &= k_\phi\alpha_\phi^l\alpha_\theta^q I_{\beta\phi} \, , \label{phasephi}
\end{align}
where the integrals are given by 
\begin{align}
I_{\alpha\theta} &= \frac{1}{2\pi\omega_\theta} \int_0^{2\pi} \cos^m\left(r\omega_\theta t + \beta_\phi\right)\cos^n\left(\omega_\theta t + \beta_\theta\right)\sin\left(\omega_\theta t + \beta_\theta\right) d\left(\omega_\theta t + \beta_\theta\right) \, , \label{I1} \\
I_{\beta\theta} &= \frac{1}{2\pi\omega_\theta} \int_0^{2\pi} \cos^m\left(r\omega_\theta t + \beta_\phi\right)\cos^n\left(\omega_\theta t + \beta_\theta\right)\cos\left(\omega_\theta t + \beta_\theta\right) d\left(\omega_\theta t + \beta_\theta\right) \, , \label{I2} \\
I_{\alpha\phi} &= \frac{1}{2\pi r\omega_\theta} \int_0^{2\pi} \cos^l\left(r\omega_\theta t + \beta_\phi\right)\cos^q\left(\omega_\theta t + \beta_\theta\right)\sin\left(r\omega_\theta t + \beta_\phi\right) d\left(r\omega_\theta t + \beta_\phi\right) \, , \label{I3} \\
I_{\beta\phi} &= \frac{1}{2\pi r\omega_\theta} \int_0^{2\pi} \cos^l\left(r\omega_\theta t + \beta_\phi\right)\cos^q\left(\omega_\theta t + \beta_\theta\right)\cos\left(r\omega_\theta t + \beta_\phi\right) d\left(r\omega_\theta t + \beta_\phi\right) \, . \label{I4} 
\end{align}

We are unaware of any standard definition for recoupling coefficients to re-express nonlinear functions of Fourier components in terms of the original components, akin to the Clebsch-Gordan machinery for spherical harmonics \cite{Edmonds60}, 
which would allow a more compact and simplified discussion of more general equations than (\ref{thetaosc}-\ref{phiosc}).
Others have created their own representations \cite{CheungZaki14}.  For now we note that these integrals (\ref{I1}-\ref{I4}) can be evaluated by rewriting the first two cosines, first in terms of exponentials and then as a multi-binomial expansion.  Thus, using expressions such as
\begin{align}
\cos^m\left(r\omega_\theta t + \beta_\phi\right)\cos^n\left(\omega_\theta t + \beta_\theta\right) = \left(\frac{1}{2}\right)^{m+n} \sum_{\mu=0}^m\sum_{\nu=0}^n \binom{m}{\mu}\binom{n}{\nu} e^{i\left[\left(m-2\mu\right)\left(r\omega_\theta t + \beta_\phi\right) + \left(n-2\nu\right)\left(\omega_\theta t + \beta_\theta\right)\right]} \, , 
\end{align}
the integrals can be evaluated directly.  They are:
\begin{align}
I_{\alpha\theta} &= -\frac{1}{2\omega_\theta} \left(\frac{1}{2}\right)^{m+n} {\sum_{\mu,\nu}}^{'} \binom{m}{\mu}\binom{n}{\nu} \sin\left[ | m-2\mu | \left(\beta_\phi-r\beta_\theta\right)\right]  \, , \\
I_{\beta\theta} &= \;\;\;\frac{1}{2\omega_\theta} \left(\frac{1}{2}\right)^{m+n} {\sum_{\mu,\nu}}^{'} \binom{m}{\mu}\binom{n}{\nu} \cos\left[\left(m-2\mu\right)\left(\beta_\phi-r\beta_\theta\right)\right] \, , \\
I_{\alpha\phi} &= \;\; \frac{1}{2r\omega_\theta} \left(\frac{1}{2}\right)^{l+q} {\sum_{\lambda,\chi}}^{'} \binom{l}{\lambda}\binom{q}{\chi} \sin\left[ | q-2\chi | \left(\frac{1}{r}\beta_\phi-\beta_\theta\right)\right]  \, , \\
I_{\beta\phi} &= \;\; \frac{1}{2r\omega_\theta} \left(\frac{1}{2}\right)^{l+q} {\sum_{\lambda,\chi}}^{'} \binom{l}{\lambda}\binom{q}{\chi} \cos\left[ \left( q-2\chi \right) \left(\frac{1}{r}\beta_\phi-\beta_\theta\right)\right]  \, ,
\end{align}
where the primed sums are only taken over indices that satisfy the following relationships:
\begin{align}
\left(m-2\mu\right)r + \left(n-2\nu\right) &= \pm 1 \, , \\
\left(l - 2\lambda\right)r + \left(q - 2\chi\right) &= \pm r \, .
\end{align}

We find that saturation fixed points of the slow equations (\ref{amptheta}-\ref{phasephi}) require $l=1$ (an unforced coupling term linear in $\phi$, so that $\alpha_\phi$ can be divided out of \eqref{ampphi}), $m\ne0$ (the unforced oscillator feeds back on the forced oscillator), $I_{\beta\phi} = 0$ (determining a phase relation between the oscillators), $I_{\alpha\phi} \ne 0$, and $\left( I_{\alpha\theta}^2 + I_{\beta\theta}^2 \right) \ne 0$.
To find explicit solutions, we combine the first two equations \eqref{amptheta} and \eqref{phasetheta} and place this alongside the third equation \eqref{ampphi}, 
\begin{align}
\frac{f^2}{4\omega_\theta^2} &=\left (-\frac{\alpha_\theta c_\theta}{2} + k_\theta \alpha_\phi^{m} \alpha_\theta^{n} I_{\alpha\theta}\right)^2 + \left(k_\theta \alpha_\phi^{m} \alpha_\theta^{n} I_{\beta\theta}\right)^2 \, , \label{combined} \\
0 &=  \alpha_\phi \left( -\frac{c_\phi}{2} + k_\phi\alpha_\phi^{l-1}\alpha_\theta^q I_{\alpha\phi} \right) \, . \label{third}
\end{align}
Aside from the trivial linear-quiescent response, there may exist a saturation solution when $l=1$,
\begin{align}
\alpha_\theta^q &= \frac{c_\phi}{2k_\phi I_{\alpha\phi}} \, , \label{alphathetasat} \\
\alpha_\phi^{m} &= \frac{1}{2k_\theta \left(I_{\alpha\theta}^2+I_{\beta\theta}^2 \right)\alpha_\theta^n} \left( \alpha_\theta c_\theta I_{\alpha\theta} \pm \sqrt{\frac{f^2}{\omega_\theta^2} \left(I_{\alpha\theta}^2 + I_{\beta\theta}^2 \right) - \alpha_\theta^2 c_\theta^2 I_{\beta\theta}^2} \, \right) \, , \label{alphaphisat}
\end{align}
which solution only exists when the forcing is great enough to ensure that the numerator in \eqref{alphaphisat} is positive and real (the correct sign choice is $+$, as $I_{\alpha\theta}\le 0$  for the cases considered here), 
and if the denominators in (\ref{alphathetasat}-\ref{alphaphisat}) involving integrals do not vanish.
For driving at the natural frequency, the appearance of a saturated solution results in the loss of stability of the trivial solution, but appropriate detuning can lead to subcritical bifurcations \cite{Nayfeh73}.

\section{Results}

We confirmed the accuracy of the theoretical predictions through numerical solution of equations (\ref{thetaosc}-\ref{phiosc}), setting $\omega_\theta = 1$ and using identical coefficients $c_\theta = c_\phi = k_\theta = k_\phi = \epsilon$.
We set a default value of $\epsilon = 0.01$, but also examined other values, and let the forcing amplitude $0 \le f \le 3\epsilon$.  Steady-state peak-to-peak amplitudes were measured after a sufficiently long time; as expected, significantly longer times were necessary near bifurcation points.  
We examined every case with integer exponents $0\leq l \leq 2$, $1\leq m \leq4$, $0\leq n \leq4$, $1\leq q \leq4$, and harmonic ratios $r=\frac{1}{2}, \frac{1}{m}, \frac{1}{q}$, and many additional cases with $l=3, 4$ and $r=1, 2, \frac{3}{2}, \frac{m}{q}, \frac{q}{m}$. 
Values of zero for $m$ or $q$ correspond to lack of information flow between oscillators, and are not of interest.
We usually observe nothing of interest for $r$ values other than $\frac{1}{2}$ and $\frac{3}{2}$, these two cases being quite similar, so in further discussion we take as default value the classic parametric resonance case $r=\frac{1}{2}$, and introduce the notation ($l$, $m$, $n$, $q$) to refer to particular oscillator pairs.

Theory and numerics agree for all cases examined, with the exception of $l=0$ cases.  
For these cases, $I_{\alpha\phi}=0$ and the averaging theory predicts the trivial  response, but the observed response involves an order $\epsilon$ response in $\phi$ at $\omega_\theta$ or $2\omega_\theta$ rather than this oscillator's natural frequency $\omega_\phi$.  This response appears to be independent of $r$, although we did not systematically check all cases.  However, as no bifurcations are observed in these cases, we will not discuss them further.

\begin{figure}[h]
	\includegraphics[width=4in]{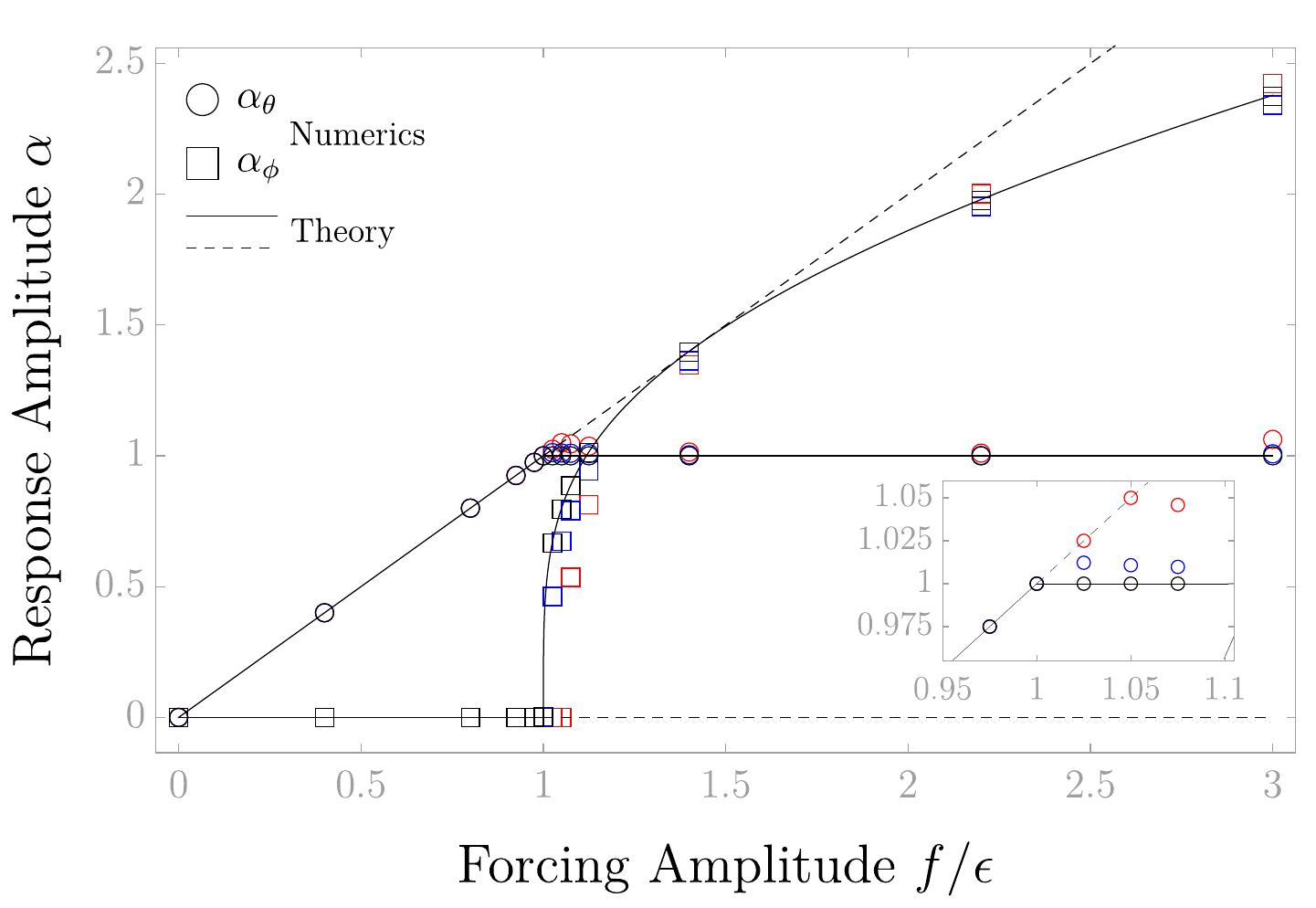}
	\captionsetup{width=4in}
	\caption{Steady-state response amplitudes as functions of forcing amplitude for the saturating case $(l, m, n, q) = (1, 2, 1, 1)$, $r = \tfrac{1}{2}$, and three values of the perturbation parameter: $\epsilon=0.01$ (black), $\epsilon=0.25$ (blue), and $\epsilon=0.5$ (red). }
	\label{satexample}
\end{figure}

Many cases of the general system show only the trivial response.
We find saturation bifurcations for oscillator pairs with $l=1$, $m=2, 4$, $n=0, 1, 2, 3, 4$, and $q=1, 3$, containing a variety of nonlinear terms, not simply quadratic as is sometimes claimed.  At a critical value of forcing, the linear-quiescent response is replaced by a saturated response, with the unforced amplitude $\alpha_\phi$ already exceeding the forced amplitude $\alpha_\theta$ once the forcing is slightly higher than critical.  Figure \ref{satexample} illustrates the saturation effect in the $(1, 2, 1, 1)$ system, and shows that the perturbative method gives quite accurate results even for moderate values of $\epsilon$.
In this example, the integrals (\ref{I1}-\ref{I4}) are $I_{\alpha\theta} = 0$, $I_{\beta\theta} = \frac{1}{4}$, $I_{\alpha\phi} = \frac{1}{2} \sin\left(2\beta_\phi - \beta_\theta\right)$, $I_{\beta\phi} = \frac{1}{2} \cos\left(2\beta_\phi - \beta_\theta\right)$.  At a fixed point, the phases are such that $2\beta_\phi - \beta_\theta = \frac{\pi}{2}$, and thus $I_{\alpha\phi} = \frac{1}{2}$ and $I_{\beta\phi} = 0$.  Inserting these and the simple choice of identical coefficients gives $\alpha_\theta = 1$ and 
$\alpha_\phi^2 = 2\sqrt{\left(f/\epsilon\right)^2-1}\,$
for the saturated solution.

\begin{figure}[h]
	\includegraphics[width=4in]{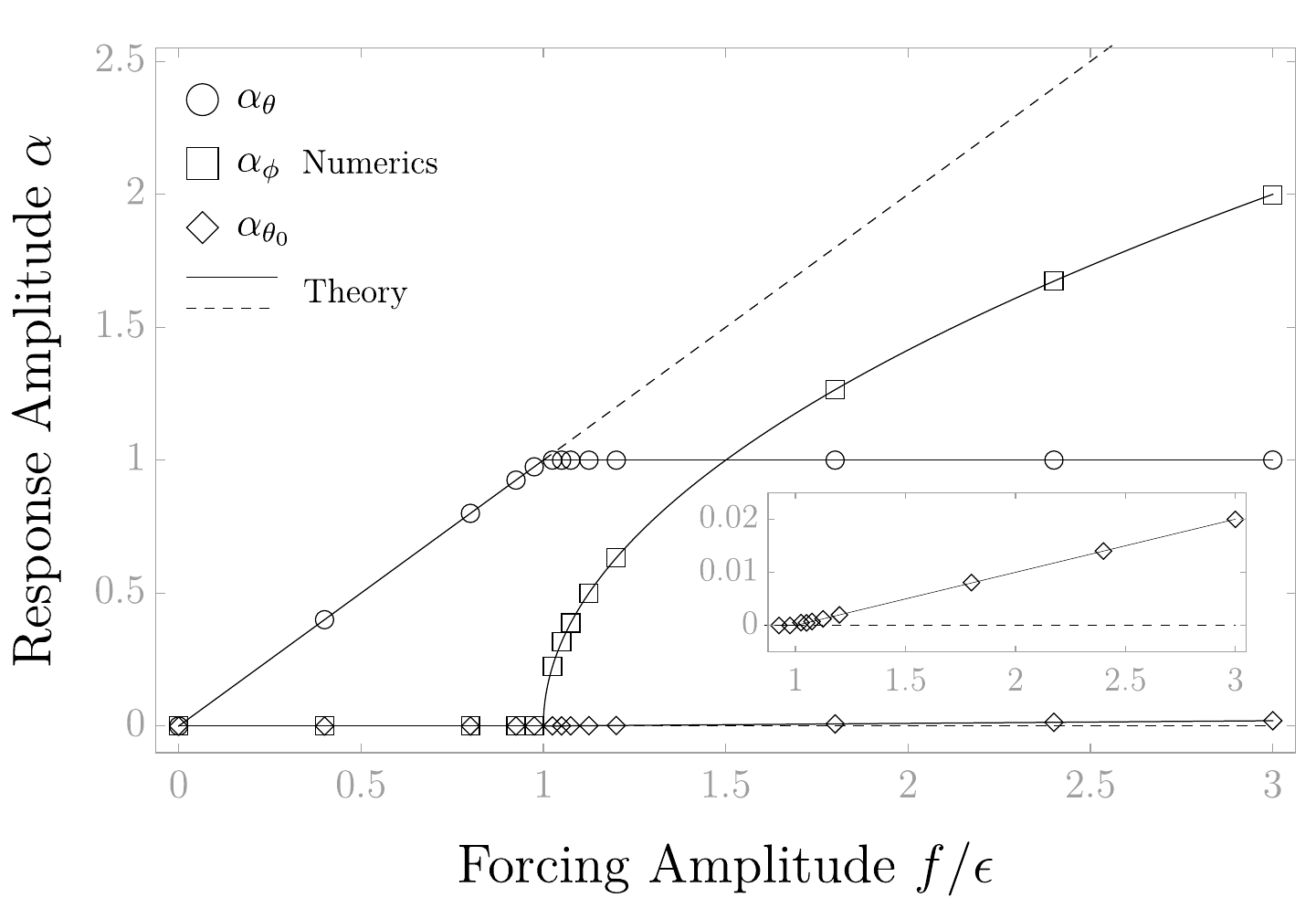}
	\captionsetup{width=4in}
	\caption{Steady-state response amplitudes as functions of forcing amplitude for the saturating case $(l, m, n, q) = (1, 2, 0, 1)$, $r = \tfrac{1}{2}$, $\epsilon=0.01$.  This system also displays a DC response, as highlighted in the inset. We define the amplitude $\alpha_{\theta_0} = |\theta_0|$;  here $\theta_0$ is a negative offset.}
	\label{n0example}
	\centering
\end{figure}

We also observe a curiosity which has not been noticed, or at least not remarked upon, in the literature, but is one that appears in the original ship motion example of saturation \cite{Nayfeh73}, as well as the spring-pendulum system \cite{LeeHsu94}.  If the integer $m+n$ is even, the saturated forced oscillator exhibits a constant shift of order $\epsilon$, which can be seen quite clearly when plotting the response as a function of time.
This can be quantified by modifying the \emph{ansatz} to include small constant terms,
$\theta = \theta_0 + \alpha_\theta\cos\left(\omega_\theta t + \beta_\theta\right)$, $\phi = \phi_0 + \alpha_\phi\cos\left(r\omega_\theta t + \beta_\phi\right)$, and 
examining the balance of restoring terms and the constant Fourier component of the nonlinear terms in (\ref{thetaosc}-\ref{phiosc}).
For the ship motion case $(1, 2, 0, 1)$, we find $\theta_0 = -\frac{k_\theta}{2\omega_\theta^2}\alpha_\phi^2$
, $\phi_0 = 0$.  
Figure \ref{n0example} shows the agreement between theory and numerics.  


A subset of the saturation fixed points subsequently undergo a Hopf bifurcation at moderately higher values of the forcing.
These cases are 
$(1, m, n, q)$ for $m=2,4$, $n=1,2$, $q=1,3$, 
$(1, 4, n, 1)$ for $n=3,4$, and
$(1, m, 4, 3)$ for $m=2,4$, all with $r = \frac{1}{2}$ or $r = \frac{3}{2}$.
Both oscillators display quasiperiodic motion, resembling what has been described for similar systems in various ways including as ``modulated'', ``almost periodic'', and an ``exchange of energy'' \cite{Sethna65, SethnaBajaj78, Nayfeh73, Yamamoto77, YamamotoYasuda77, Miles84, Mook85, Mook86, Streit88, NayfehZavodney88, BalachandranNayfeh91, Oh00, Tondl97}.
A numerical Fourier analysis reveals the appearance of second frequencies close to the natural frequencies of each oscillator, leading to beat frequencies that scale as $f_B \sim \epsilon$.

\begin{figure}[h]
	\includegraphics[width=6.5in]{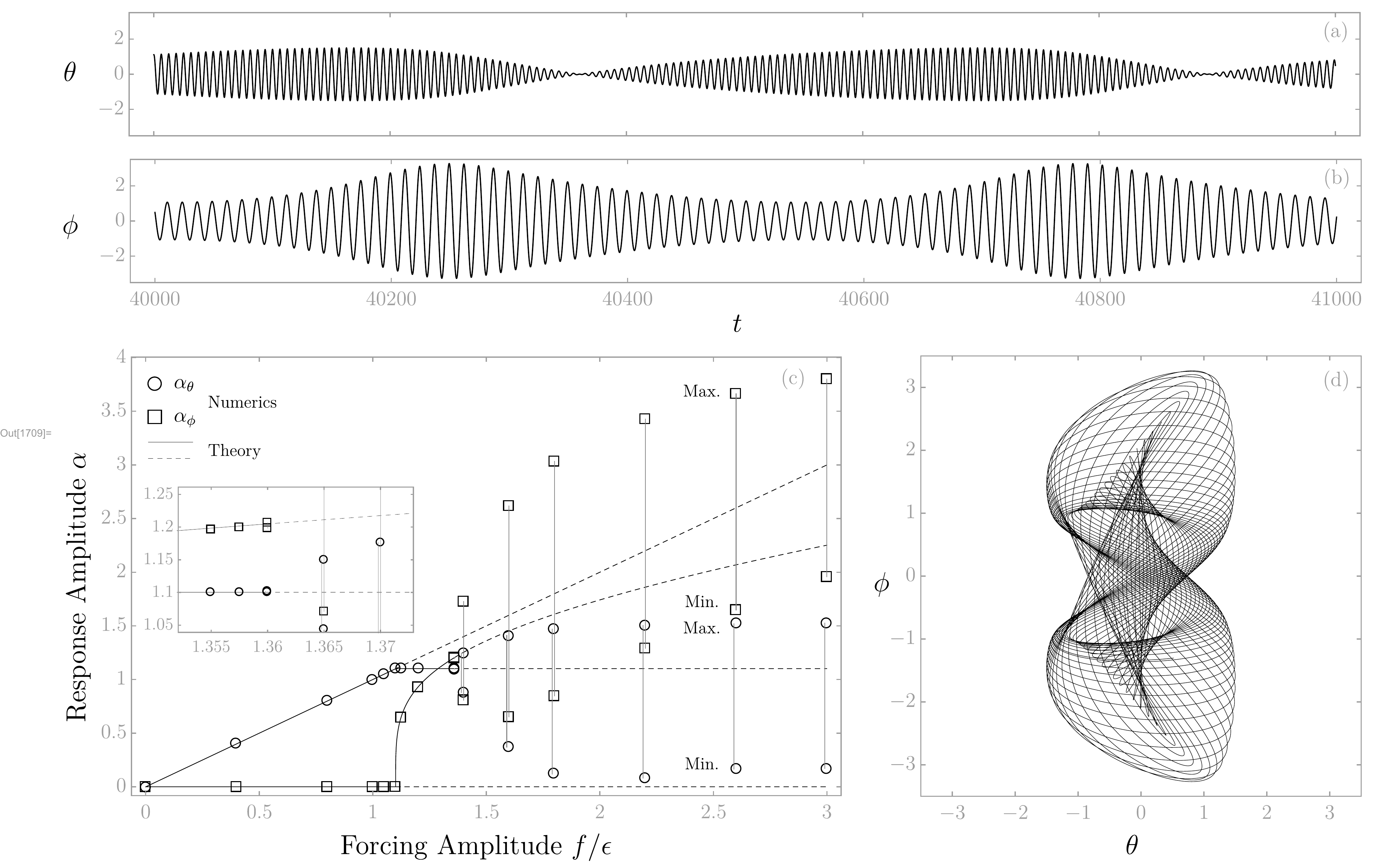}
	\captionsetup{width=6.5in}
	\caption{The case $(l, m, n, q) = (1, 2, 1, 3)$, $r = \tfrac{1}{2}$, $\epsilon=0.01$ displays saturation followed by a Hopf bifurcation to quasiperiodicity.  Response of (a) forced and (b) unforced oscillators at $f=2\epsilon$; (c) steady-state response amplitudes as functions of forcing amplitude, including the maximum and minimum amplitudes after the Hopf bifurcation; (d) parametric plot at $f=2\epsilon$.}
	\label{oscfig}
\end{figure}

We can detect the loss of stability of the saturated state by linearizing the slow equations (\ref{amptheta}-\ref{phasephi}) around the saturation fixed point.
For example, consider the case (1, 2, 1, 3).  Expressing the fixed point phases from the integrals in terms of the fixed point amplitudes, we obtain the linearized form
\begin{equation}
\begin{pmatrix} \dot\alpha_\theta^{(1)} \\ \dot\beta_\theta^{(1)} \\ \dot\alpha_\phi^{(1)} \\ \dot\beta_\phi^{(1)}\end{pmatrix} = \begin{pmatrix} -\frac{c_\theta}{2} & -\frac{k_\theta}{4}{\alpha_\phi^{(0)}}^2\alpha_\theta^{(0)} & 0 & 0 \\ \frac{k_\theta}{4}{\alpha_\phi^{(0)}}^2 & -\frac{c_\theta}{2} & \frac{k_\theta}{2}\alpha_\phi^{(0)} & 0 \\ \frac{3}{2}\frac{c_\phi \alpha_\phi^{(0)}}{\alpha_\theta^{(0)}} & 0 & 0 & 0 \\ 0 & \frac{c_\phi}{2} & 0 & -c_\phi \end{pmatrix} \begin{pmatrix} \alpha_\theta^{(1)} \\ \beta_\theta^{(1)} \\ \alpha_\phi^{(1)} \\ \beta_\phi^{(1)} \end{pmatrix}. \label{stabilitymat}
\end{equation}
The forcing $f$ appears implicitly through the fixed point amplitudes in the Jacobian.  Inserting the simple choice of identical coefficients, we find that a pair of eigenvalues become unstable at $f \approx 1.36\epsilon$.
Figure \ref{oscfig} shows the behavior of this system, and its agreement with the stability analysis.

\begin{figure}[h]
	\includegraphics[width=6.5in]{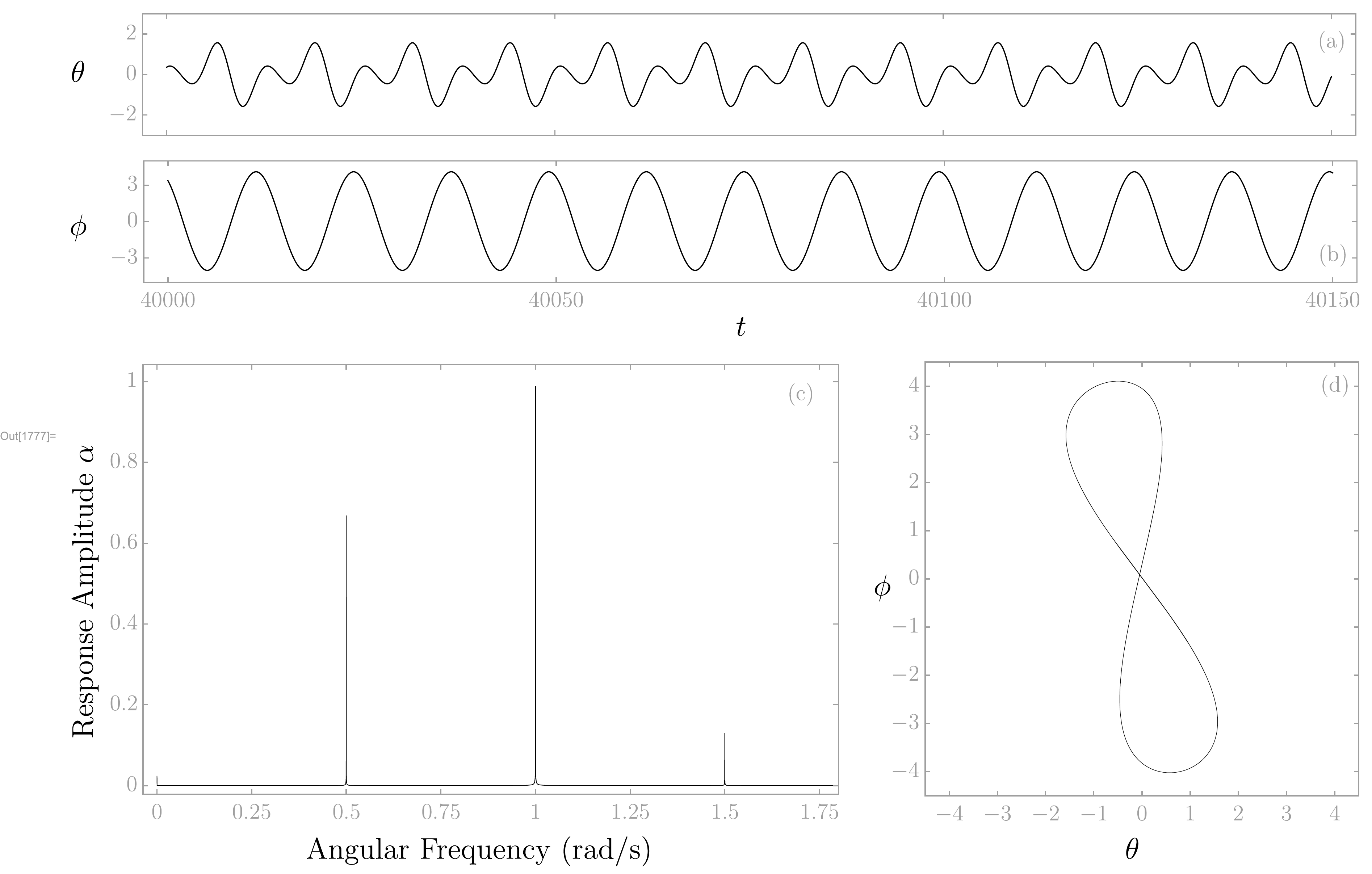}
	\captionsetup{width=6.5in}
	\caption{The multifrequency case $(l, m, n, q) = (1, 3, 0, 1)$, $r = \tfrac{1}{2}$, $\epsilon=0.01$, at $f=2\epsilon$.  Response of (a) forced and (b) unforced oscillators; (c)  frequency spectrum of the forced oscillator; (d) parametric plot.}
	\label{multifig}
	\centering
\end{figure}

Finally, we observe another remarkable group of oscillator pairs that do not saturate, but instead undergo a bifurcation to periodic behavior with multiple commensurate frequencies.  
These cases are
$(1, 1, 0, 1)$,
$(1, 1, n, q)$ for $n=1,2,3,4$, $q=1,3$, and
$(1, 3, n, q)$ for $n=0,2,4$, $q=1,3$, all with $r = \frac{1}{2}$.  Similar behavior was also observed for many of these cases with $r = \frac{3}{2}$, but this was not systematically documented.
Constant shifts 
are also observed in both forced and unforced oscillators for some cases, but this does not follow the same pattern of dependence on the exponents as in the saturation cases.  Some of these systems also display subsequent Hopf bifurcations to quasiperiodicity, and some appear to become more seriously unstable, at moderate values of forcing.
As an example, we show the response of the $(1, 3, 0, 1)$ system in Figure \ref{multifig}.  At a critical value of forcing, the linear-quiescent response is replaced by three frequencies in the forced oscillator (its natural frequency $\omega_\theta$, along with $\frac{1}{2}\omega_\theta$, and $\frac{3}{2}\omega_\theta$), and 
one frequency in the unforced oscillator (its natural frequency $\frac{1}{2}\omega_\theta$), along with small constant shifts in both.

As yet, we have not found any documentation of this multi-frequency bifurcation in the closely related literature, and we do not have a criterion to predict multi-frequency states.  Even ignoring the DC response, the system requires eight equations for amplitudes and phases; attempting to employ the method of averaging by ignoring the higher harmonic and using six equations does not lead to a solution.

In conclusion, our perturbative analyses predicting saturation bifurcations and the stability and response of the saturated state, including zero-frequency offsets, are in excellent agreement with numerical results for a class of weakly nonlinear coupled oscillator pairs.  We also observed another type of bifurcation to a multi-frequency periodic state that we believe merits further study.

\section*{Statement of author contribution}
AR conceived the study and theoretical approach, and performed initial theoretical calculations and numerics.  JAH modified and extended the theoretical approach.  HGW performed theoretical calculations and numerics, and made the figures.  HGW and JAH wrote the paper.

\section*{Acknowledgments}
We thank S S Gupta for alerting us to some references pertaining to vibration localization.

\bibliographystyle{unsrt}


\end{document}